\documentclass[twocolumn,showpacs,aps,floatfix]{revtex4} 
\usepackage{graphicx} 
\usepackage{bm} 
\usepackage{amsmath} 
\newcommand{\ignore}[1]{} 
 
\def\pr{\prime}

\def\half{{1 \over 2}} 
\begin{document} 
%
%
\title{ Biperiodic superlattices and the transparent state} %
\author{D. W. L. Sprung$^a$, L. W. A. Vanderspek$^b$, W. van 
Dijk$^{a,b}$, J. Martorell$^{c}$ and C. Pacher$^{d}$ } 
%
\affiliation{ $^a$Department of Physics and Astronomy, McMaster 
University\\   Hamilton, Ontario L8S 4M1 Canada
}
\affiliation{
$^b$Department of Physics, Redeemer University College\\
  Ancaster, Ontario L9K 1J4 Canada
}
\affiliation{
$^c$Departament
  d'Estructura i Constituents de la Materia, Facultat F\'{\i}sica,\\
   University of Barcelona, Barcelona 08028, Spain
}
\affiliation{ 
$^d$Austrian Research Centers GmbH - ARC, 
Smart Systems Division, 
Donau-City-Str. 1, 1220 Vienna, Austria 
}
\date{\today}
%
\begin{abstract}
We study biperiodic semiconductor superlattices, 
which consist of alternating cell types, one with wide wells and 
the other narrow wells, separated by equal strength barriers. 
If the wells were identical, it would be a simply periodic system 
of $N = 2n$ half-cells. When asymmetry is introduced, an allowed band 
splits at the Bragg point into two disjoint allowed bands. The 
Bragg resonance turns into a transparent state located close to 
the band edge of the lower(upper) band when the first(second) well 
is the wider. Analysis of this system gives insight into how band 
splitting occurs. Further we consider semi-periodic systems having 
$N= 2n+1$ half-cells. Surprisingly these have very different 
transmission properties, with an envelope of transmission maxima 
that crosses the envelope of minima at the transparent point. 
 \end{abstract}
%
\maketitle
\section{Introduction}
\begin{figure}[htb]                     
\begin{center}
\includegraphics[width=8cm]{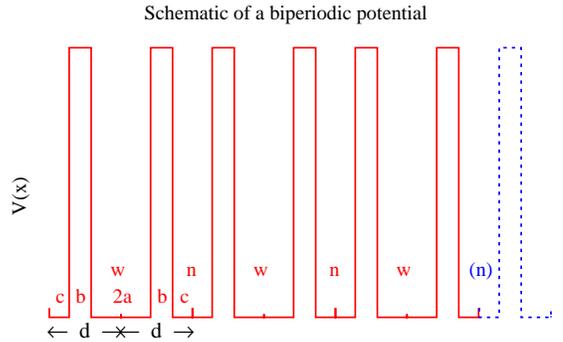} 
\end{center}
\caption{(Colour online)
Solid (red) line: Biperiodic array [1, 2] of three 
double-cells, each of width $2d$; short-dashed (blue) line an 
additional half-cell could be added at right, creating an additional 
narrow well as discussed in Sec. IV.} 
 \label{fig01} 
\end{figure}

Coquelin {\it et al}. \cite{Coq1,Coq2} carried out
experiments on electron transmission 
through a finite biperiodic GaAs/AlGaAs superlattice consisting of 
alternating types of unit cells.  Biperiodic systems occur naturally 
in crystals and polymers \cite{AC04}, but in layered semiconductor 
heterostructures one has control over the properties of the cells. 
As illustrated by the red (solid) line in Fig. \ref{fig01}, 
Coquelin's system had identical barriers of width $b = 3.8$ nm, while 
there are two alternating well widths, $2a = 4.3$ nm (wide) and $2c = 
3.8$ nm (narrow), which changes it from a simply 
periodic to a biperiodic system. 

We will consider the basic unit, called a half-cell, to comprise 
three layers: two well segments of GaAs having widths $a,\, c$ 
separated by an AlGaAs barrier of width $b$. A full or ``double-cell" 
consists of a half-cell plus another which is its mirror image. The 
double-cell of width $2d = 2(a+b+c)$ centered on the origin, is marked 
in Fig. \ref{fig01} at the left of the three-cell array, in red. It has 
layers of widths $c,\, b,\, 2a,\, b,\, c$ and is overall reflection 
symmetric about its mid-point. Due to the two barriers enclosing a 
well of width $2a$, there will be quasi-bound states which 
show up as resonances in electron scattering from the double cell. 
When $N$ such cells are juxtaposed, the wells will be alternately 
wide and narrow; hence the name biperiodic array. Each double cell is 
a symmetric cell, and the complete array has reflection symmetry. By 
exchanging the values of $a$ and $c$, the character of the array will 
change from say $w, n, w, \cdots$ to $n, w, n, \cdots$, where $w$ and 
$n$ stand for wide and narrow wells respectively.  

   The experimental device \cite{Coq1,Coq2} had $n=3$ double cells, 
so there were $2n-1 = 5$ wells enclosed between barriers. This number 
was chosen because it provides enough separation between the 
measured transmission resonances for them to be distinguished. 
When $a=c$, the simply periodic system of $N = 2n$ symmetric cells is 
known 
\cite{FPP,CP81,PC82,VC86,YKT89,KL92,KL95,BDS96,Gom99,Yang00,GS01,SYR02,PP05,GK03,YMS02,MYS02,SD02} 
to exhibit $N-1$ transmission resonances in each allowed band, according to 
 \begin{eqnarray}
|t_N|^{2} &=& [ 1 + \sinh^2 \mu\,  \sin^2 N\phi_h  ]^{-1} \ge \cosh^{-2} \mu~, 
\label{eq:bip01}
\end{eqnarray}
where $\phi_h$ is the Bloch phase of the half-cell, and 
$\mu$ is its impedance parameter in the Kard parameterization Eq. 
(\ref{eq:bip10}) of the transfer matrix\cite{SMM03,SMM04}. 
The fundamental band structure depends on those properties of the 
half-cell, while the width and spacing of individual resonances 
depends on the number $N$ of half-cells, via  
$N\phi_h = m\pi$, with $m = 1,\, 2,\, ... N-1$. 

When $a \ne c$, the half-cell is asymmetric under reflection, though 
the double cell remains symmetric. Each allowed band develops a band 
gap near the Bragg point $\phi_h = \pi/2$. An allowed band of the 
$n=3$ double-cell system should show two resonances. As is seen in 
Fig. \ref{fig02}, one of the allowed bands contains a third resonance 
which we will identify as a transparent state at which the 
impedance parameter $\mu \to 0$, causing the envelope of minimum 
transmission probability to be pushed up to unity. Such a state 
occurs at a fixed energy, independent of the number of double 
cells included, as can be seen from eq. \ref{eq:bip01}. This 
transparent state lies very close to a band edge, in 
the lower (upper) split band when the wide (narrow) well is first in 
line for incident electrons. 

The purpose of this paper is to explain why and how the transparent 
state arises when the half-cell becomes asymmetric, and why it is 
located very close to the split band edge.  In Sec. IV we will 
consider the surprising effect of including an additional half cell, 
as suggested by the dashed blue line at the right in Fig. \ref{fig01}.

\section{Transfer matrix analysis}
\subsection{General}
We will follow the line of argument of Shockley \cite{Surf}, who 
studied surface states of a finite static periodic potential 
whose unit cell is symmetric about its mid-point. We extend his 
method to allow for the position and energy-dependent effective mass, 
required in a semiconductor superlattice. Since the full cell has reflection 
symmetry, it is sufficient to solve the Schr\"odinger equation 
for the half-cell $0 < x < d$. The even ($g(x)$) and odd ($u(x)$) parity 
solutions take the boundary values $g(0)=1, g'(0)=0$ and $u(0)=0, 
u'(0)=1$ at the origin. [Note: Following the development of Appendix 
A, the prime means take the derivative, and then divide by the 
variable factor $m m^*/\hbar$ to allow for the effective mass, $m^* 
\sim 0.07$, which is dimensionless.] The transfer matrix for the 
half-cell is 
 \begin{eqnarray}
W_R &=& W_{0,d} = \begin{pmatrix}g & u \\ g' & u'\end{pmatrix}~, 
\quad {\rm and} \nonumber \\ 
W_L &=& W_{0,-d}^{-1}  = \begin{pmatrix}u' & u \\ g' & g \end{pmatrix}~. 
\label{eq:bip02}
\end{eqnarray}
where the wave functions without argument are always evaluated at the 
point $x=d$. Placing the half-cells in reverse order, simply 
interchanges the elements $g$ and $u'$.  

Since the Wronskian of two solutions is a constant, $\det W = g u' - 
g' u = 1$. In dealing with a constant potential and constant $m^*$, 
for example, one has 
 \begin{eqnarray}
g(x) = \cos q x \quad u(x) = (m m^*/ \hbar q) \,  \sin q x
\label{eq:bip03}
\end{eqnarray}
which are chosen so that at $q=0$ (well bottom), we have 
solutions $g(x)=1$ and $u(x)= m m^*\, x/\hbar $. 

For the symmetric double cell $-d < x < d$, the transfer matrix is 
 \begin{eqnarray}
W = W_{-d,d} &=& W_R \, W_L = 
\begin{pmatrix}g & u \\ g' & u' \end{pmatrix} \, 
\begin{pmatrix}u' & u \\ g' & g \end{pmatrix} \nonumber 
\\ 
&=& \begin{pmatrix}gu' + g' u & 2ug\\ 2u'g' & gu'+g'u   \end{pmatrix} \, .
\label{eq:bip04}
\end{eqnarray}
To transform to the ingoing/outgoing waves representation we proceed 
as in Appendix A with $\nu = \hbar k/(m m^*)$ as the 
velocity outside the potential. Unlike $W$, 
the matrix $M_{-d,d}$ operates from right to left. For the full cell it 
is given by eq. \ref{eq:a08}. 

\begin{figure}[htb]                     
\begin{center}
\begin{tabular}{cc} 
a) & \includegraphics[width=8cm]{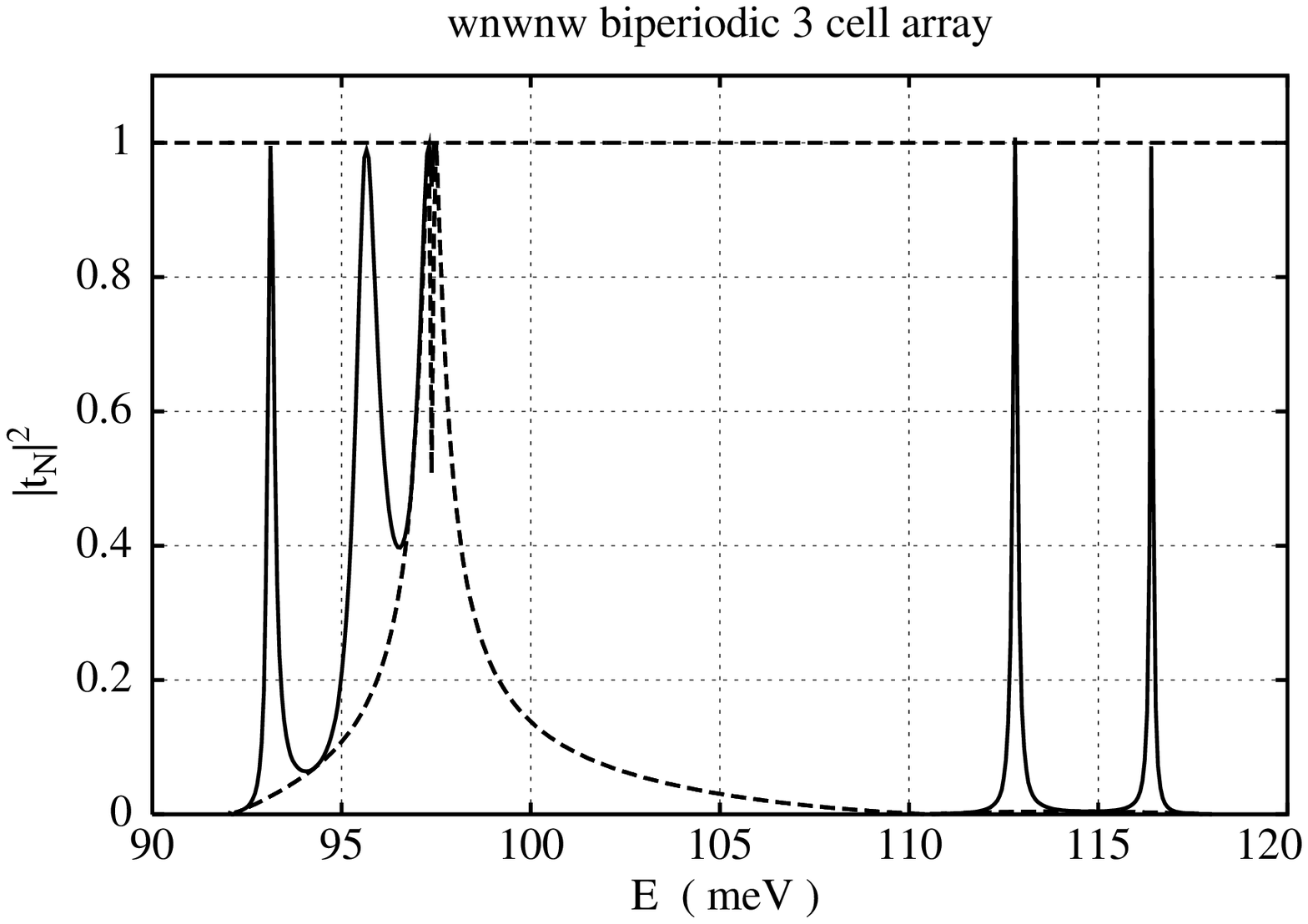} \\ 
b) & \includegraphics[width=8cm]{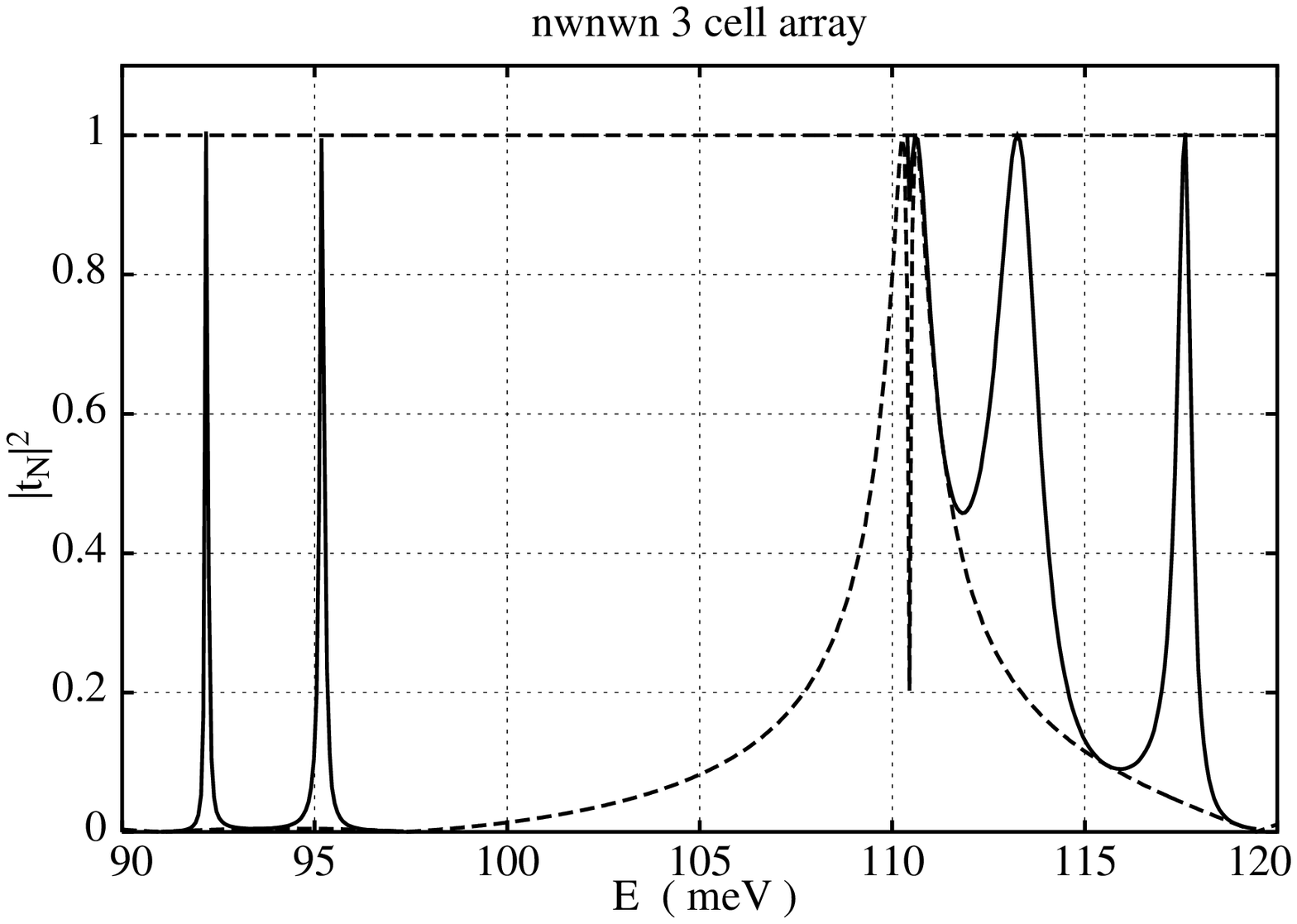} \\ 
\end{tabular}
\end{center}
\caption{Transmission (solid line) for the 3-cell biperiodic array
of Fig. 1:  
(a) wide well first; and (b) narrow well first. The dashed line is 
the envelope of transmission minima in the allowed zones and an upper 
bound in the band gap. }  
 \label{fig02} 
\end{figure}

Shockley's first step was to derive the Bloch phase $\phi$ of the 
{\it double} cell in terms of the solutions of the half cell, by 
appealing to Floquet's theorem. He obtained 
 \begin{eqnarray}
\tan^2 \phi/2 = - \frac{\gamma}{\lambda} = - \frac{g'u}{gu'}~,
\label{eq:bip05}
\end{eqnarray}
where $\gamma, \lambda$ are the log-derivatives of the half-cell 
solutions $g, u$ respectively.  One finds easily that 
 \begin{eqnarray}
\cos \phi = \frac{1- \tan^2 \phi/2}{1+ \tan^2 \phi/2}  = 
 \frac{gu' + ug' }{gu' - ug'} \quad , 
\label{eq:bip06}
\end{eqnarray}
where the denominator is $\det W = 1$. We write 
$W$ in the parameterized form 
 \begin{eqnarray}
W &=& \begin{pmatrix}\cos \phi & (1/Z) \sin \phi \\ -Z \sin \phi & \cos 
\phi \end{pmatrix} \quad {\rm with} \nonumber \\ 
Z^2 &=& - {\gamma \lambda}~. 
\label{eq:bip07} 
\end{eqnarray}
The log-derivatives $\gamma,\, \lambda$ determine the 
effective velocity $Z$ and the Bloch phase $\phi$ at each energy, for 
a symmetric double cell. Again, in case of a constant potential across the 
cell, they would be 
 \begin{eqnarray}
\gamma &=& -(\hbar q/m m^*) \tan qd\, ;  \qquad  
\lambda = (\hbar q/m m^*) \cot qd \nonumber \\
Z &=& (\hbar q/m m^*)\ ;\qquad \quad  \tan^2 \phi/2 = \tan^2 qd~. 
\label{eq:bip08} 
\end{eqnarray}
It is evident that in an allowed band $\gamma$ and $\lambda$ must 
have opposite signs, to make both $\phi$ and $Z$ real. Conversely in 
a forbidden band $\gamma$ and $\lambda$ have the same sign. 

Inserting eq. \ref{eq:bip07} into 
eq. \ref{eq:a08}, gives $M = M_{-d,d} = $
 \begin{equation}
 \begin{pmatrix} \cos \phi -i \frac{\sin \phi}{2} ( \nu/Z +Z/\nu)
                               & i \frac{\sin \phi}{2}  (\nu/Z-Z/\nu ) \\
 -i \frac{\sin \phi}{2}  (\nu/Z-Z/\nu ) & 
\cos \phi +i \frac{\sin \phi}{2} (\nu/Z+Z/\nu) \end{pmatrix}  
\label{eq:bip09}
\end{equation} 
Defining $e^\mu = \nu/Z$ as the ratio of velocities outside/inside the 
potential region, we obtain the Kard parameterization in our standard 
form \cite{SMM03,SMM04}: 
 \begin{eqnarray}  
\begin{pmatrix} \cos \phi -i \sin \phi\, \cosh \mu 
                               & i\, \sin \phi\, \sinh \mu \\
 -i\, \sin \phi\, \sinh \mu & 
\cos \phi +i\, \sin \phi\, \cosh \mu \end{pmatrix}
\label{eq:bip10} 
\end{eqnarray}
We call $\mu$ the impedance parameter, since a slower velocity 
inside the cell corresponds to a greater impedance. 

\subsection{Asymmetric delta-barrier cells} 
To gain insight into how band splitting occurs, we consider a simple 
model which replaces the square barrier cells illustrated in Fig. 
\ref{fig01} by delta-function barriers of strength $\Omega d = 1.403 
\pi$, (a value chosen to give results similar to those of the 
Coquelin potential \cite{Coq1}). In the well sections, we have $\nu = 
\hbar k/ m m^*$ (see Eqs. (A3, A5)).The transfer matrix for a half-cell 
of width $d = a+c$ is 
 \begin{eqnarray}
W_R &=& W_c \, W_\delta \, W_a  =   
 \begin{pmatrix}\cos kc & (\sin kc) / \nu \\ -\nu \sin kc & \cos kc \end{pmatrix} 
\, \times       \nonumber \\     
&& \qquad \begin{pmatrix} 1 & 0 \\ 2\Omega\,\nu/k & 1 \end{pmatrix} \, 
    \begin{pmatrix}\cos ka & (\sin ka) / \nu \\ -\nu \sin ka & \cos ka \end{pmatrix} \,  \nonumber \\ 
&=& \begin{pmatrix} g & u \\ g' & u' \end{pmatrix} \qquad {\rm with} \nonumber \\ 
g &=& \cos kd + \frac{\Omega d}{kd} (\sin kd - \sin ks) \nonumber \\ 
g' &=& -\nu \left[ \sin kd - \frac{\Omega d}{kd} (\cos kd + \cos ks) 
\right] \nonumber \\ 
u &=& \frac{1}{\nu} \left[ \sin kd - \frac{\Omega d}{kd} 
(\cos kd - \cos ks) \right] \nonumber \\ 
u' &=& \cos kd + \frac{\Omega d}{kd} (\sin kd + \sin ks)~. 
\label{eq:bip11} 
\end{eqnarray} 
We have written $s = a-c$; $s/d$ is the asymmetry. The double-cell 
has two barriers and a well of width $2a$ between. Reversing the sign 
of $s$ interchanges $a$ and $c$, which is equivalent to putting the 
two half-cells in the opposite order. When $s >0$ a 
wide well occurs on both ends of the biperiodic superlattice. 

The band structure depends only on the location of the zeroes and 
poles of $\gamma, \, \lambda$. These locations do not change if the 
$\nu$ in front of the off-diagonal elements $g^\pr$ and $u$ is 
multiplied by a constant factor. To reduce the number of parameters 
in play, we replace {\it that} $\nu \to kd$, equivalent to saying that 
$\hbar/m m^* = \nu/k = d$. Then the transfer matrix of the 
delta-barrier model is a function of dimensionless 
variables $kd$, $s/d = ks/kd$ and $\Omega d$. The figures are drawn 
as functions of $kd$; the lowest allowed band ends when $kd \approx 
\pi$.

\begin{figure}[htb]                     
\begin{center}
\includegraphics[width=8cm]{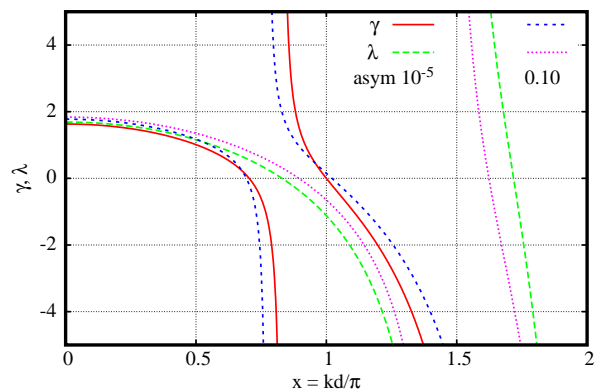} 
\end{center}
\caption{(Colour online) 
Log-derivatives $\gamma, \, \lambda$ of the solutions 
$g(kd)$ and $u(kd)$ for the delta-barrier system, for  
very small $10^{-5}$ and moderate 0.10 asymmetry; 
$\gamma$ is shown as solid (red) line and short-dashed (blue) line; 
$\lambda$ as long-dashed (green) and dotted (mauve) lines, 
respectively.}  
 \label{fig03} 
\end{figure} 
\begin{figure}[htb]                     
\begin{center}
\includegraphics[width=8cm]{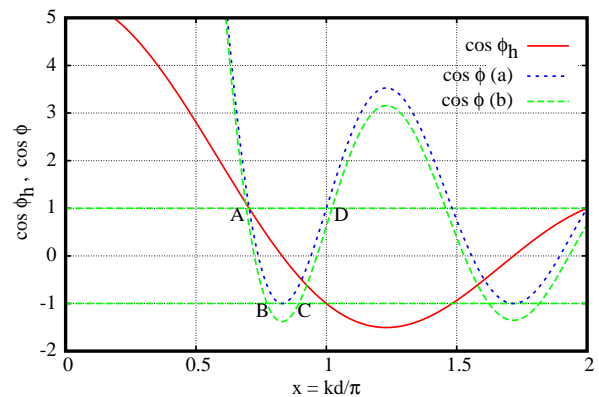} 
\end{center}
\caption{(Colour online) 
$\cos \phi_h$ of the half-cell (solid/red line), 
and $\cos \phi$ of the double-cell, for 
the delta-barrier system; (a) for very small ($10^{-5}$) asymmetry, 
(long-dash/green line), and (b) for moderate ($0.10$) asymmetry 
(short-dash/blue line). }
 \label{fig04} 
\end{figure}

  Figure \ref{fig03} shows the evolution of the log-derivatives 
$\gamma$ and $\lambda$ versus $kd$, for two values of the asymmetry 
parameter, $s/d = 10^{-5}$, and $0.10$. For a fixed cell width $d$, 
the plot shows the energy dependence (via $kd$). Larger $s/d$ moves 
the poles to the left, seen by the red and blue lines for $\gamma$. 
For $\lambda$ (green and mauve lines) the node shifts to the right. 
Allowed bands occur when $\gamma$ and $\lambda$ have opposite signs, 
which covers much of the interval for $kd$ between $0.7 \pi$ and 
$\pi$. For the $s = 10^{-5}$ asymmetry, the pole of $\gamma$ at 
approximately $0.83 \pi$ almost coincides with a node of $\lambda$, 
creating an infinitesimal forbidden band there. For a reflection 
symmetric half-cell, they would exactly coincide, and cancel, giving 
a vanishing gap. For the larger asymmetry, the separation between the 
pole and node increases, widening the gap. A magnified view of this 
region is shown in Fig. \ref{fig06}.

In Fig. \ref{fig04} we see the corresponding Bloch phases. The red 
(solid) lines are the $\cos \phi_h$ of the half cell, while the blue 
(dotted) and green (dashed) lines are the $\cos \phi$ of the double 
cell. For line (a) the angle $\phi$ is almost equal to $2 \phi_h$, so 
when $\phi_h = \pi/2$, the dashed line scarcely descends below $-1$. 
However, for moderate asymmetry (line(b)), the undershoot is evident, and a 
sizeable band gap opens up between $kd = 0.77$ (B) and $0.89 \pi$ 
(C). The outer band edges, marked by $A$ and $D$, shift outwards a 
little at the same time. 

\begin{figure}[htb]                     
\begin{center}
\begin{tabular}{cc} 
a) & \includegraphics[width=8cm]{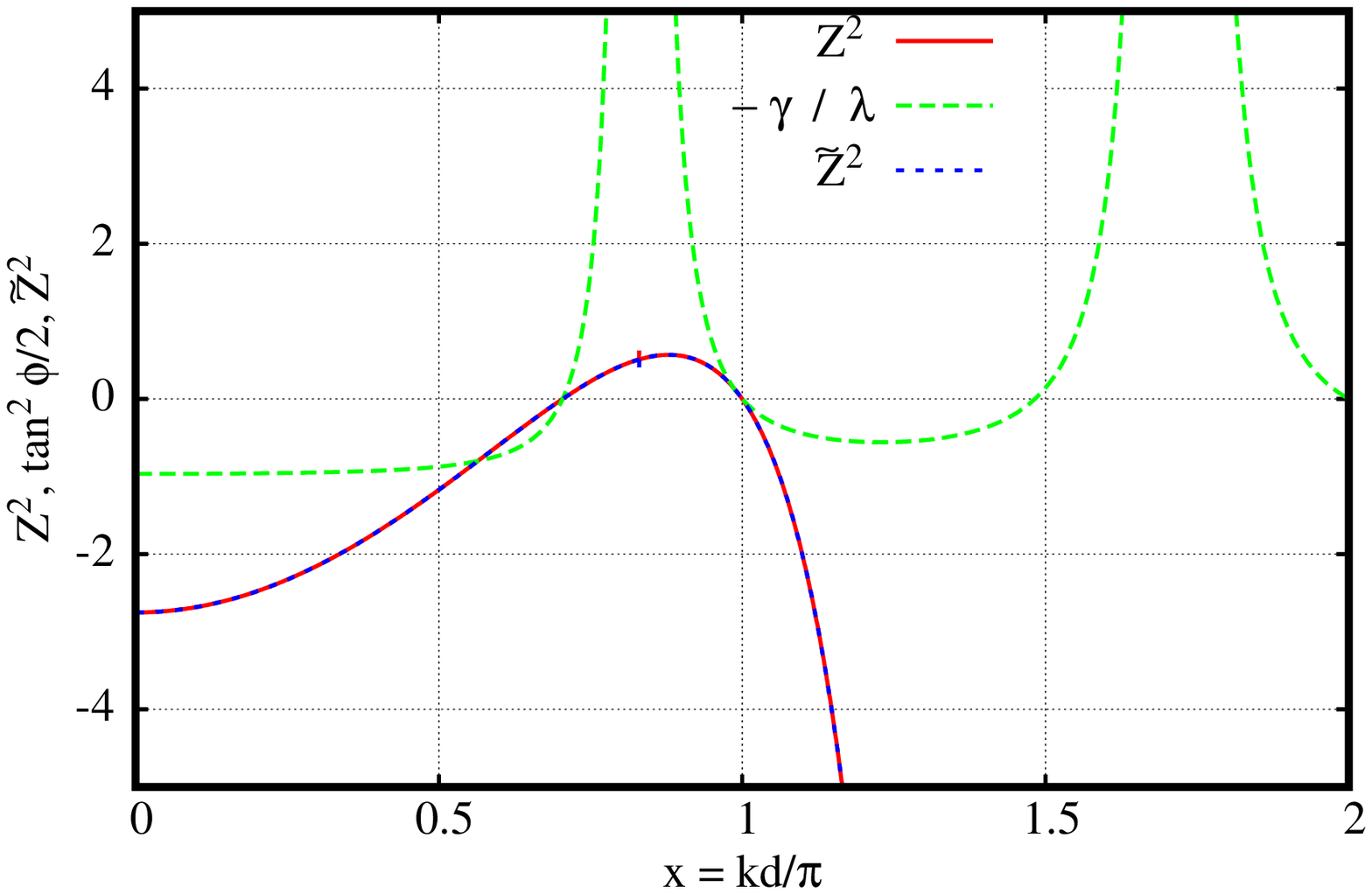} \\ 
b) & \includegraphics[width=8cm]{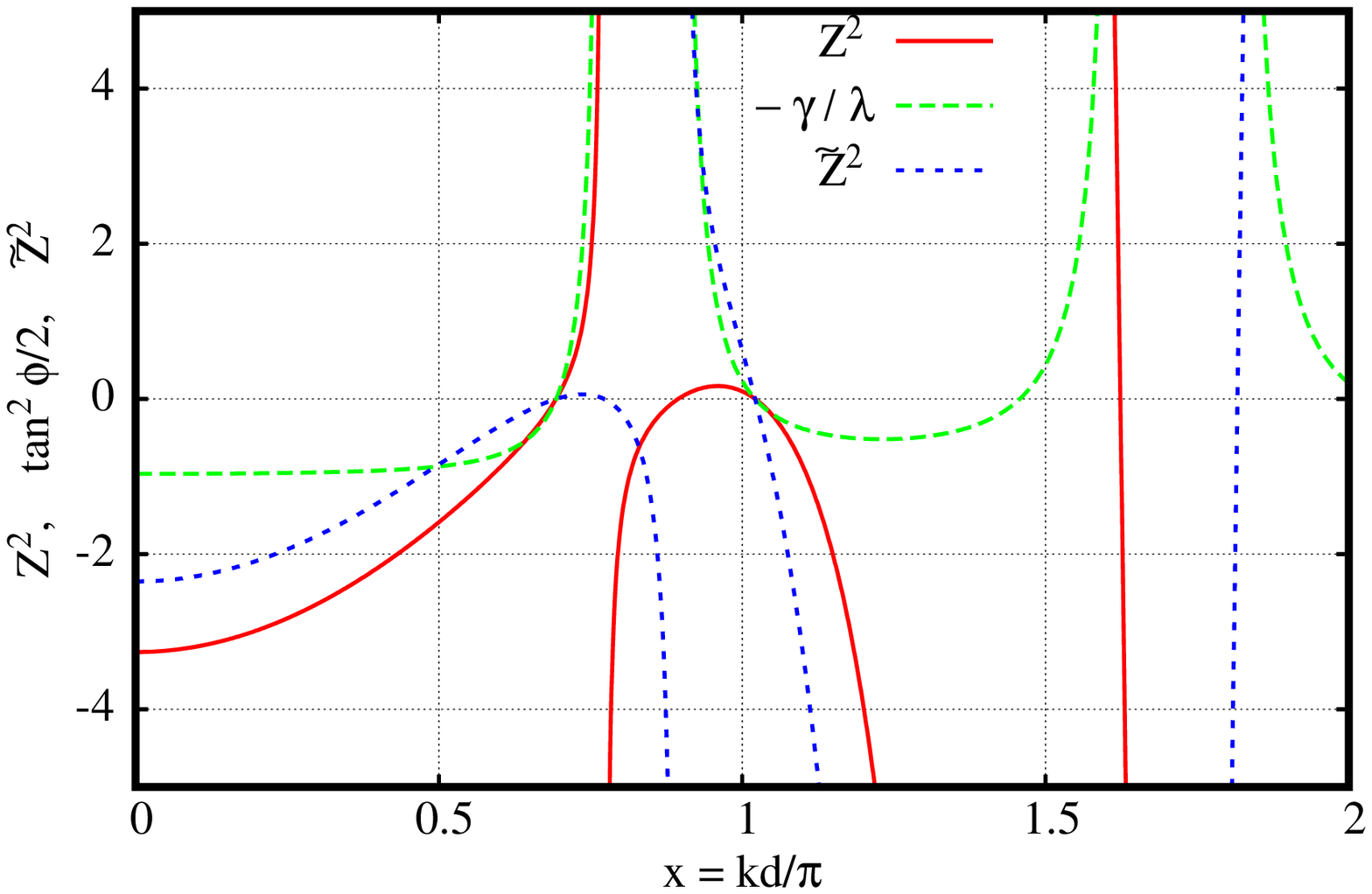} \\ 
\end{tabular}
\end{center}
\caption{(Colour online) 
$\tan^2 \phi/2$: long-dashed (green) line; $Z^2$: solid (red) line; 
and $\tilde{Z}^2$: short-dash (blue) line,  for the delta-barrier 
system; (a) case of very small ($10^{-5}$) asymmetry; (b) moderate 
($0.10$) asymmetry. In (a) the two poles almost coincide and 
cancel, leaving only a small glitch.}  
 \label{fig05} 
\end{figure}

\begin{figure}[htb]                     
\begin{center}
\begin{tabular}{cc} 
a) & \includegraphics[width=8cm]{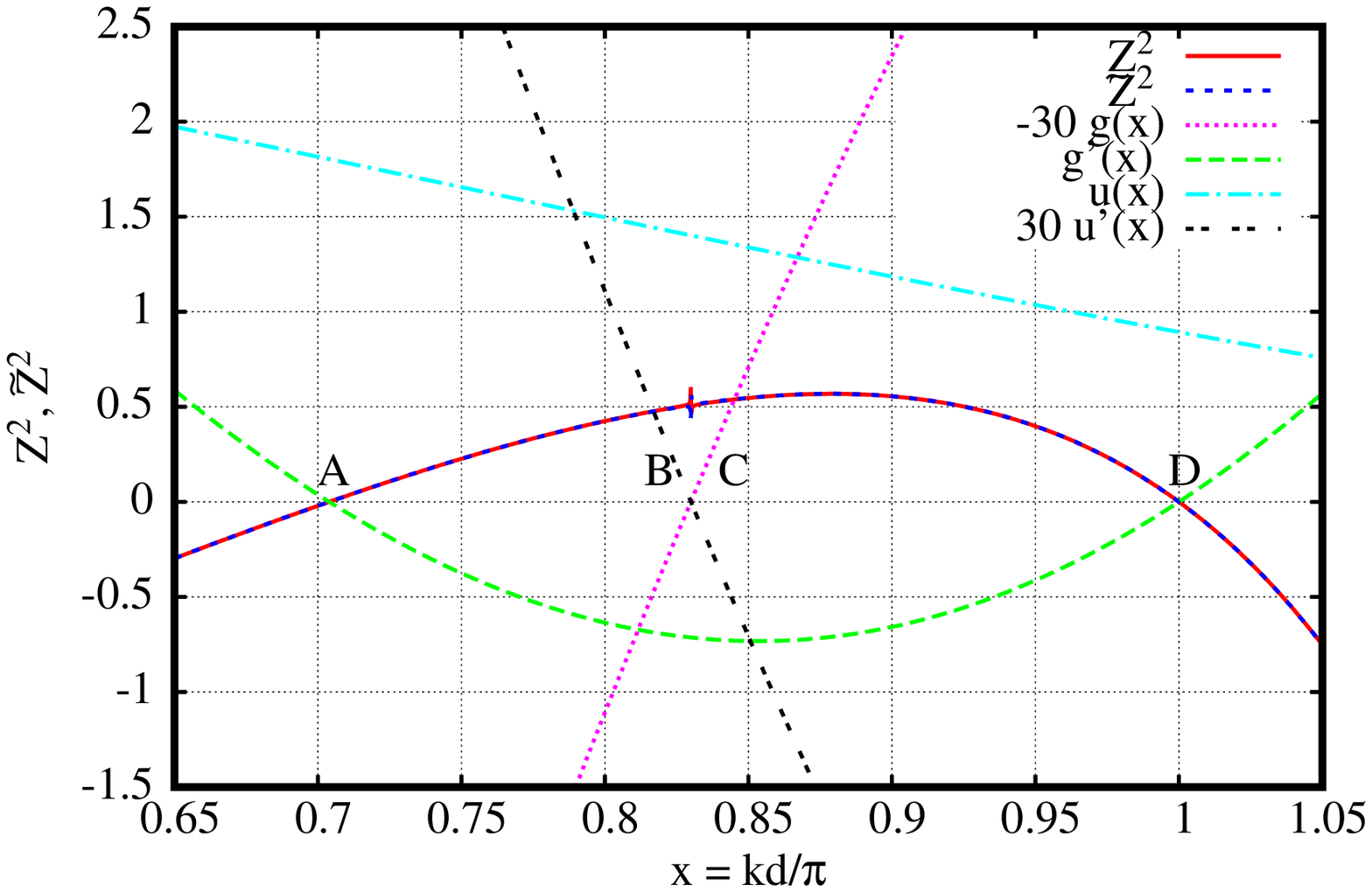} \\ 
b) & \includegraphics[width=8cm]{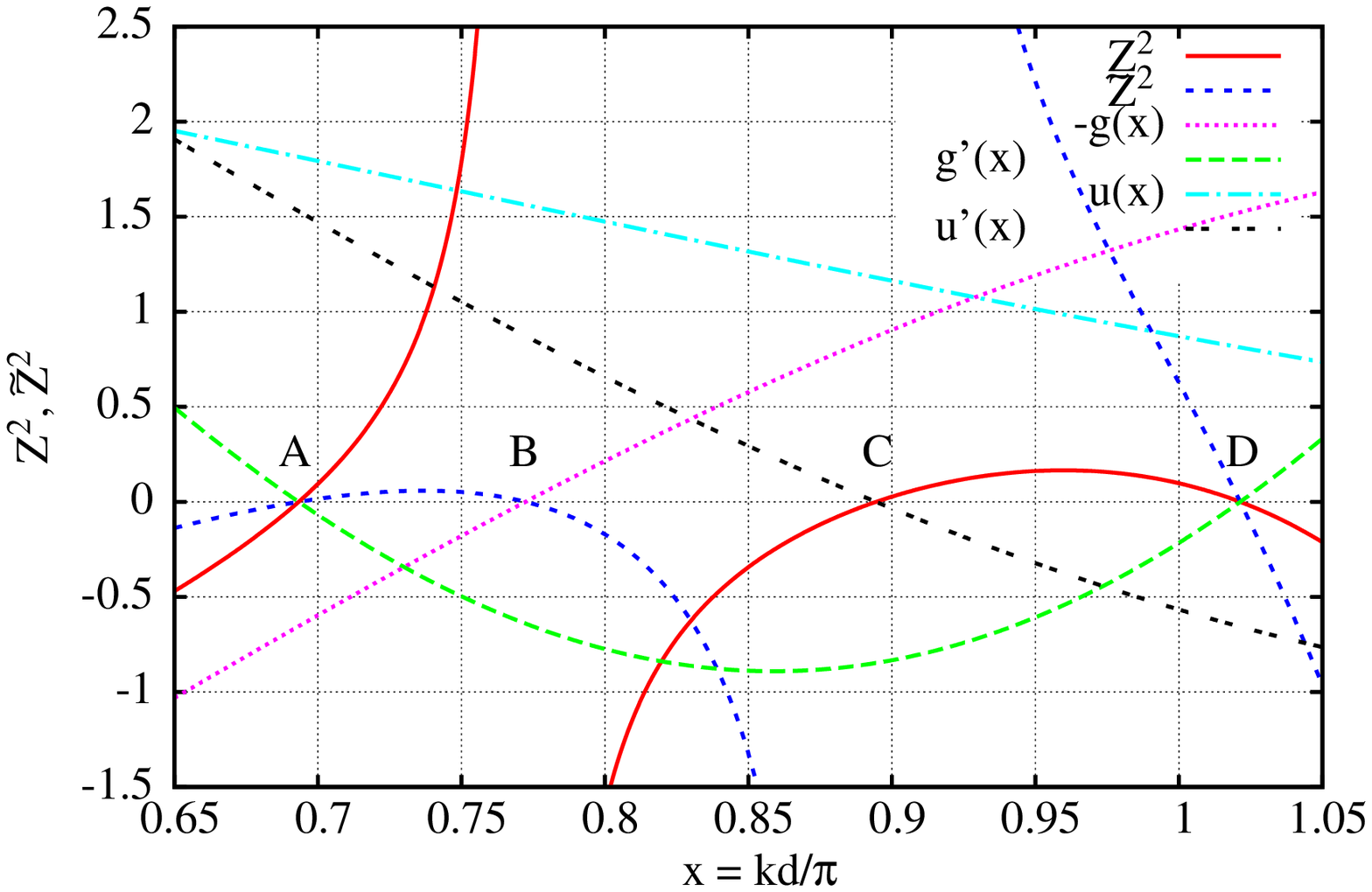} \\ 
\end{tabular}
\end{center}
\caption{(Colour online) 
Pole region of $Z^2$ and $\tilde{Z}^2$, and wave functions $g(x),\, g'(x)$ 
and $u(x),\, u'(x)$; 
(a) case of $10^{-5}$ asymmetry; 
(b) moderate ($0.10$) asymmetry. Lines identified at upper right. }    
 \label{fig06} 
\end{figure}

In Fig. \ref{fig05}, the green (long dashed) line is $-\gamma/\lambda = 
\tan^2 \phi/2$; positive values are necessary for an allowed 
band to exist. The solid line is $Z^2$ defined in eq. 
\ref{eq:bip07}. The short dashed line is the alternative value 
$\tilde{Z}^2 = - g^\pr u^\pr / (g u)$, which results when we 
interchange the values of $a$ and $c$, choosing the opposite asymmetry. 
Since this changes the sign of $s$, it is equivalent to interchanging 
the values of $g$ and $u'$, which leaves the trace of $W_R$ and 
therefore $\phi_h$ unchanged.  In panel (a) $\tilde{Z}^2$ lies almost 
on top of $Z^2$, because the asymmetry is practically zero; only a 
small glitch (due to finite steps in drawing) marks the location on 
the curve near $0.83 \pi$. In panel (b) the poles of $Z^2$ and 
$\tilde{Z}^2$ separate cleanly. When $a > c$, $Z^2$ is large below 
the band gap, and small above the band gap, which runs from $kd = 
0.77$ to $0.89 \pi$. When $a < c$, $\tilde{Z}^2$ applies and those 
properties reverse. 

The two panels of Fig. \ref{fig06} provide a magnified view of the 
split band region. The straight mauve (dotted) and black (double-dash) 
lines are multiples of $g$ and $-u^\pr$, the negative sign imposed so 
that their crossing point can be easily identified. This is the point 
at which Tr$W_R = 0$, which makes $\phi_h = \pi/2$. For a symmetric 
half-cell, this is the energy where the pole and node coincide, 
and cancel each other.  The point labelled B is a node of $g$ and a 
pole of $\gamma = g^\prime/g$, at the lower edge of the band gap. To 
the left of the pole, $\gamma$ (red, solid) diverges, and so does 
$Z \to \infty$. Since $Z$ rises though all positive values 
between threshold at A and the pole at B, at some point it 
must equal the external velocity $\nu$, which makes the impedance 
parameter $\mu = \log{\nu/Z} \to 0$. That defines the transparent 
point, at which $|t_N|^{-2} = 1 + \sinh^2 \mu \sin^2 N\phi = 1$ 
independent of the number of cells or the value of the Bloch phase. 
This accounts for the third transmission resonance  in the lower 
allowed band of Fig. \ref{fig02}(a). Conversely, if we take the 
opposite asymmetry, (so the narrow well is first in line), then it is 
the alternate function $\tilde{Z}^2$ which applies. This pole has the 
opposite sign residue, so it is the divergence of $\tilde{Z}^2$ near 
point labelled C at the lower edge of the upper allowed band, which 
produces the transparent state. 

In Fig. \ref{fig06}(a), the two poles are very close together at BC. In the 
limit of exact reflection symmetry, they would coincide, and 
their sum would be zero. The band gap would disappear, and $\cos \phi$ 
would touch $-1$ without crossing below that line. 

\section{ General case}
To discuss the general case of barrier-type cells arranged from left 
to right in order $L R L R \cdots L/R$, we write the transfer 
matrices (in an allowed band) in terms of three real parameters as 
follows: 
 \begin{eqnarray}
W_R &=& \begin{pmatrix} g & u \\ g' & u' \end{pmatrix} \equiv 
\begin{pmatrix}e^{-\alpha} \cos \beta & 
(1/z) \sin \beta \\ -z \sin \beta & e^{\alpha} \cos \beta \end{pmatrix} 
\nonumber \\ 
W_L &=& \begin{pmatrix} u' & u \\ g' & g \end{pmatrix} 
\equiv \begin{pmatrix}e^{\alpha} \cos \beta & 
(1/z) \sin \beta \\ -z \sin \beta & e^{-\alpha} \cos \beta \end{pmatrix} 
\nonumber \\ 
W &=& W_R\, W_L = \begin{pmatrix}  \cos 2\beta & (e^{-\alpha} /z) \sin 2\beta \\ 
- z e^{\alpha} \sin  2\beta &  \cos 2\beta \end{pmatrix} 
\label{eq:bip12} 
\end{eqnarray} 

Comparing to $W$ of eq. \ref{eq:bip07}, we see that $\beta = \phi/2$, half 
the Bloch phase of the symmetric double cell, while $Z = z e^\alpha$ 
is the corresponding velocity parameter. Interchanging the half-cells 
is equivalent to reversing the sign of $\alpha$, which measures the 
degree of asymmetry of the half-cell, but leaving $\beta$ and 
$z$ unchanged. Incrementing the number of half-cells leads to the 
following rule: 
 \begin{eqnarray}
W^{(3)} &=& W_L\, W = \begin{pmatrix} e^{\alpha}\,\cos 3\beta & (1 /z) \sin 3\beta \\ 
- z \sin  3\beta &  e^{-\alpha} \cos 3\beta  \end{pmatrix} 
\nonumber \\ 
W^{(4)} &=& W^2 = \begin{pmatrix}  \cos 4\beta & (e^{-\alpha} /z) \sin 4\beta \\ 
- z e^{\alpha} \sin  4\beta &  \cos 4\beta \end{pmatrix} ~\cdots
\label{eq:bip13} 
\end{eqnarray}

\noindent
The number $(3)$ can be replaced by any odd integer, and $(4)$ by any 
even integer. 
When the index is even the system has reflection symmetry and $\phi_N 
= 2n \beta$, but $e^\alpha$ appears on the off-diagonal elements making 
$Z_N = z e^\alpha$. 
For odd orders $N = 2n+1$, the transfer matrix is modelled 
on the half-cell, (here $W_L$), which is repeated one extra time. Then 
 \begin{eqnarray}
\cos \phi_N &=& \cosh \alpha \,\, \cos (2n+1)\beta \nonumber \\ 
Z_N &=& z~. 
\label{eq:bip14} 
\end{eqnarray} 
Superlattices with an odd number of half-cells are biperiodic but not 
reflection symmetric. They exist in two forms depending on the sign of 
$\alpha$, (or what is the same thing, whether the first well on the 
left is of wide or narrow type). We will discuss their surprising 
properties below; for now we concentrate on the case $N = 2n$ which 
are symmetric biperiodic systems involving $n$ double-cells. 

\begin{figure}[htb]                     
\begin{center}
\begin{tabular}{cc} 
a) & \includegraphics[width=8cm]{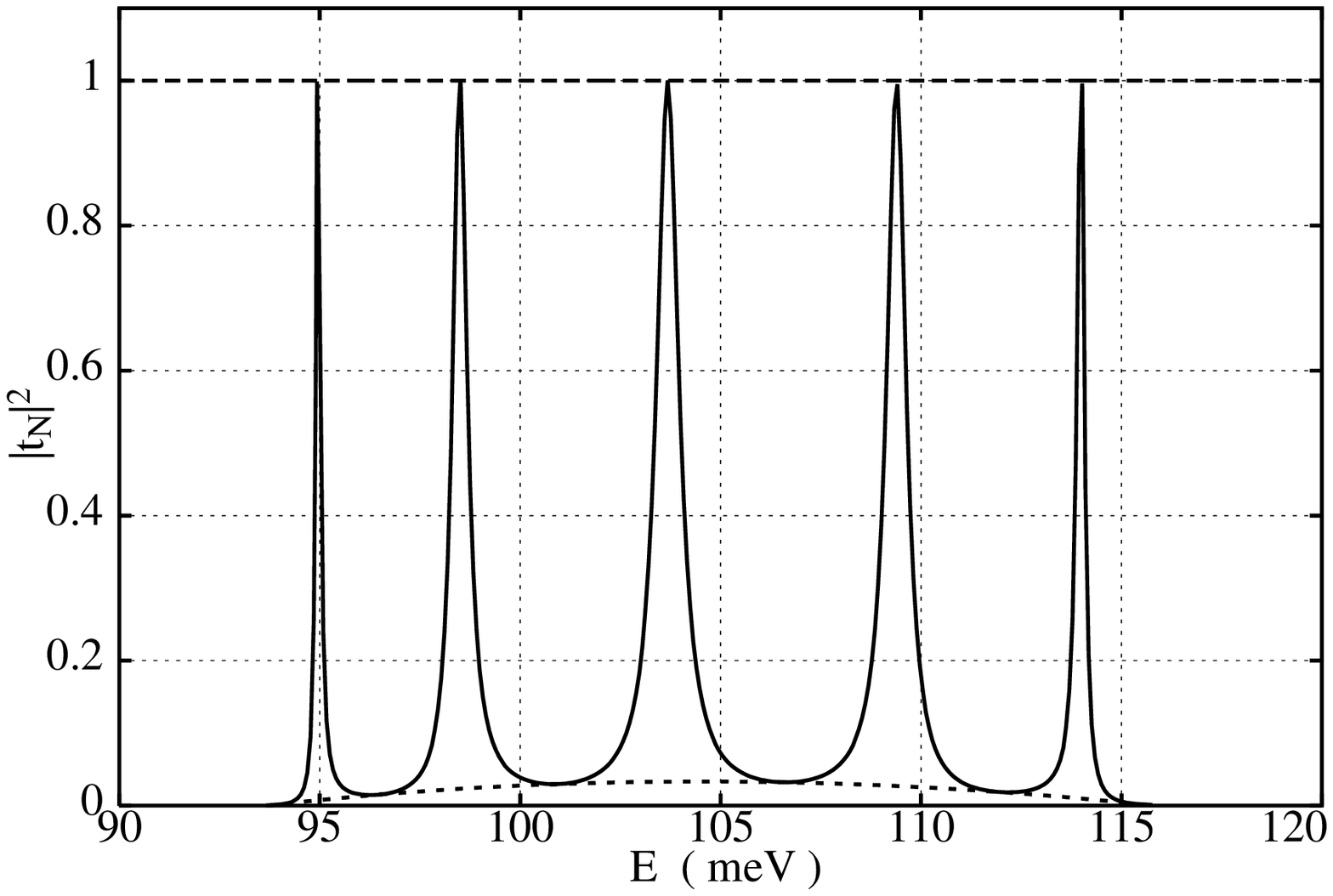} \\ 
b) & \includegraphics[width=8cm]{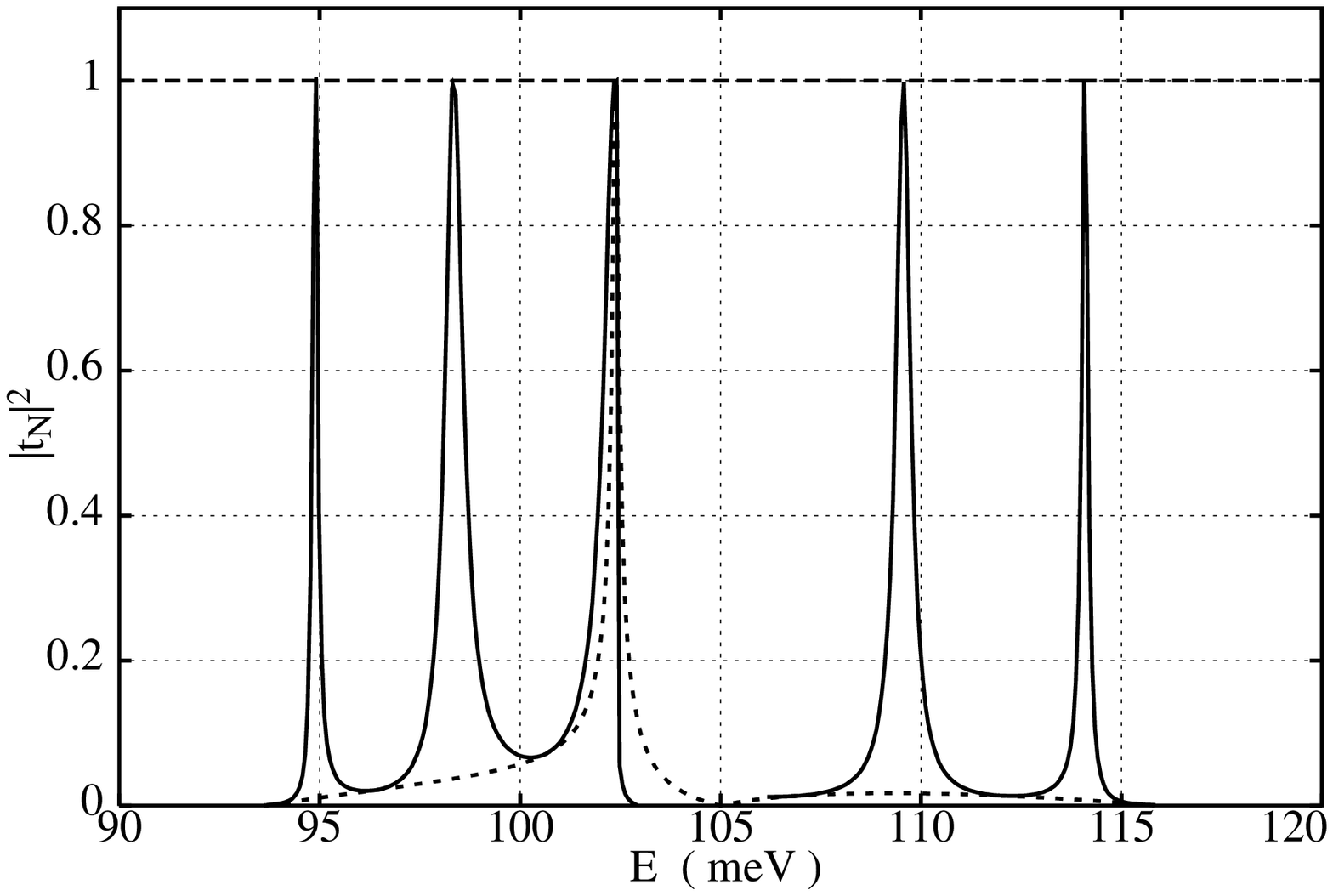} \\ 
c) & \includegraphics[width=8cm]{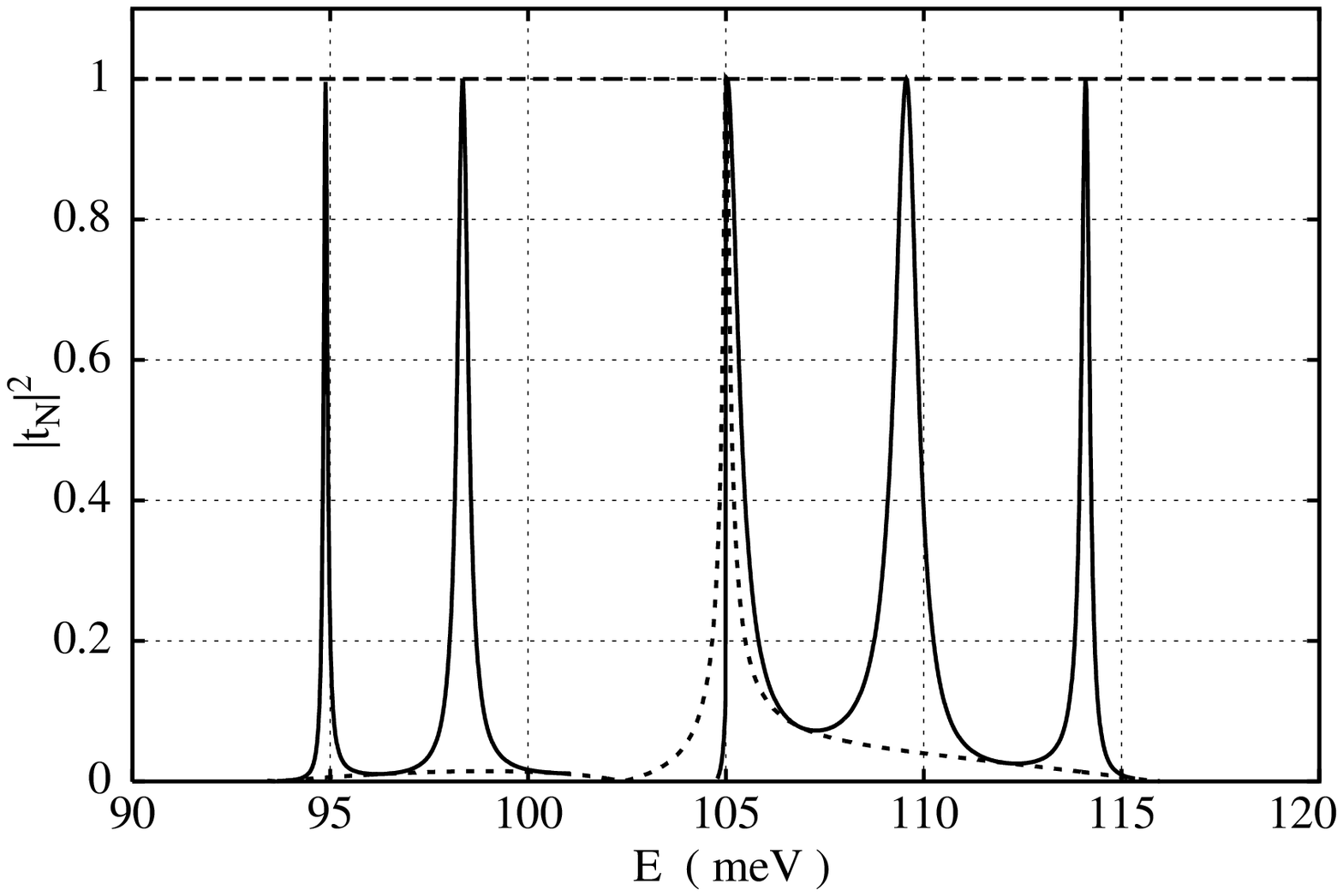} \\ 
\end{tabular}
\end{center}
\caption{Transmission of AlGaAs-barrier 3-cell biperiodic arrays: (a) zero asymmetry; (b) $2(a-c) = 0.1$ nm; (c) $2(a-c) = -0.1$ nm. Lines as in 
Fig. \protect{\ref{fig02}}.} 
 \label{fig07} 
\end{figure}

\section{Transmission in symmetric biperiodic systems} 
The transmission probabilities shown in Fig. \ref{fig02} were 
calculated for the system described in the first paper of Coquelin 
{\it et al}. \cite{Coq1}.  We took into account the variable effective 
mass and other material properties as in \cite{PG03,MPG03,PBG05}. 
Specifically, the barrier height is 288.09 meV, and the effective 
masses $m^*$ are approximately 0.074 (well) and 0.080 (barrier). 

Results shown in Figs. \ref{fig02} (a) and (b) correspond to the 
two devices with `wnwnw' and `nwnwn' well arrays. The total width of 
the allowed bands is the same for both cases because the Bloch phase 
of the double cell is the same for both orderings\cite{PG03}. 
What is different between the two cases is the impedance parameter 
$\mu_{wn} \ne \mu_{nw}$; this accounts for the different results 
obtained. In Fig. \ref{fig02}(a) the lower allowed band 
runs from 92 to 97.5 meV. The long dashed line is the envelope of 
minimum transmission, the curve $1/\cosh^2 \mu$: 
see eq. \ref{eq:bip01}. The presence of the 
transparent state with $\mu = 0$ just inside the allowed band,  
pushes the envelope of minima up to unity. This is the cause of the 
third resonance sitting close to the band edge; 
its width is driven by the envelope of minima. The envelope (dashed 
line) is continued across the forbidden band, (where it has other 
significance,) but it is seen to decay slowly across the forbidden 
zone, and is very low in the second allowed band. In panel (b), the 
order of the half-cells is reversed, and the extra state occurs in 
the upper band. 

It is impressive that even a small departure from strict periodicity has 
such a large effect on the band structure, opening a sizeable gap 
from 98. to 112. meV,  while narrowing the allowed band width as 
compared to the case of all wide or all narrow wells. The difference 
between the 3.8 and 4.3 nm widths is only 13\%. Since $\cos \phi$ and 
$\mu$ are single-cell properties, increasing the number of cells has 
no effect on the band structure; it simply squeezes more resonances 
into the bands. In a forbidden zone, $\phi \to p\pi + i\theta$. Since 
$\cosh 3\theta$ increases rapidly above 98 meV, the transmission does 
cut off sharply in the forbidden zone, even for just three 
double-cells. 

\begin{figure}[htb]                     
\begin{center}
\includegraphics[width=8cm]{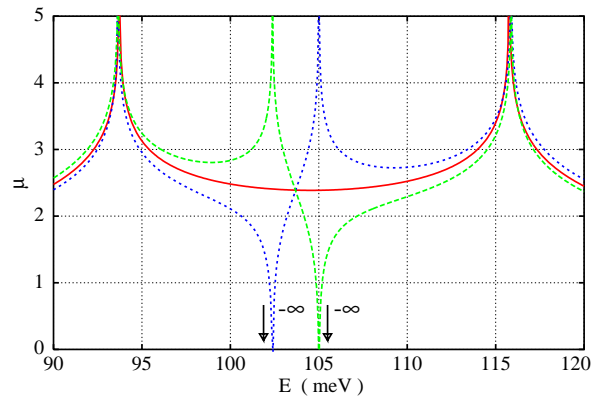} 
\end{center}
\caption{(Colour online) 
Impedance parameter $\mu$ of the 3-cell arrays of Fig. 7: 
solid (red) line, zero asymmetry; short-dashed (blue) line, small asymmetry; 
long-dashed (green) line, opposite asymmetry. The arrows imply  
divergence to $-\infty$. } 
 \label{fig08} 
\end{figure}

Further insight is gained by looking at a sequence of models very 
close to the symmetric ($a = c$) limit. In Fig. \ref{fig07}(a) we 
start with a simply periodic system using wells of the average width 
4.05 nm.  Then we made a small excursion into asymmetry by using $2a 
= 4.1$ and $2c = 4.0$ nm for the wide and narrow wells respectively. 
In Fig. \ref{fig07}(b) it is the $wnwnw$ configuration and in panel 
(c) the $nwnwn$. The gap induced is about 2.6 meV, while the outer 
band edges are little shifted.  

The transparent state clearly develops from the Bragg state of the 
symmetric double-cell, at $E_B = 104.7$ meV. The double cell has 
strong barriers, so the energies of its quasibound states $\sim 1/L^2$ 
are determined primarily by the well width \cite{SJW00}. When 
asymmetry is introduced, $2L= d \pm s$, in the notation of eq. 
\ref{eq:bip11}. The energy difference of the two cases, wide or 
narrow well, is therefore of order $ (s/d) E_B \sim 0.125 E_B = 13$ 
meV, in Fig. \ref{fig02}.  This agrees well with the locations of the 
transparent states at 97.4 and 110.5 meV, when $2d = 8.1$ nm and $2s 
= 0.5$ nm. In Fig. \ref{fig07}, $2s = 0.1$ nm and the splitting is 
2.6 meV. In a tight-binding model, the transparent state is an edge 
state of one of the split bands, in which only the wide or only the 
narrow wells are occupied at resonance, which is consistent with the 
quasibound state picture. 

In Fig. \ref{fig08}, the impedance parameters $\mu$ of the three 
situations are drawn. The red (solid) line exhibits typical behaviour 
\cite{SMM04} with a divergence of $\mu$ at the band edges. In blue 
(dotted) the band splitting has caused $\mu$ to descend steeply 
through zero at the transparent state, before diverging to $-\infty$ 
at the band gap edge, $102.5$ meV. In the forbidden band, $\mu \to 
\xi + i\pi/2$; $\xi$ rises from $-\infty$ and diverges to $+\infty$ 
at the upper edge of the band gap. The behaviour in the upper allowed 
band reverts to the typical one. In green (dashed), the negative 
divergence of $\mu$ is transferred to the lower edge of the upper 
band. The subsequent rise in the  allowed band produces the 
transparent state and the consequent large values of the envelope of 
minimum transmission, just above that band edge. 

In ref. \cite{SMM04} we interpreted the transfer matrix as a mapping 
of the system point in the complex plane, or as a hyperbolic rotation 
of a Dirac spinor around a fixed axis whose orientation is determined 
by $\mu$. The axis passes through a fixed point at distance $\tanh 
\mu/2$ from the origin. In the first allowed band of a symmetric 
system, as a function of energy, the fixed points start and end at 
$z=+1$ at the band edges, moving on the real axis. Then in the 
forbidden band, two fixed points move around the unit circle in 
complex conjugate positions, from $+1$ to $-1$. In the ensuing 
allowed band the motion starts and ends at $z = -1$. The motion which 
arises here is quite different, in that (in Fig. \ref{fig08}, blue 
line) the interior fixed point begins at $+1$ at the lower edge of 
the first allowed band, but does not return to $+1$ at the band edge; 
rather it remains on the real axis and moves through the origin to 
reach $-1$ at the lower edge of the gap. Its passage through the 
origin produces the transparent point. In the band gap, the fixed 
points do move on the unit circle, initially very quickly to reach 
$\pm i$ and then more slowly reaching $+1$ at the lower edge of the 
second allowed band. The fixed points then move on the real axis in 
the normal manner, starting and ending at $+1$ in the upper split 
band. The fixed points must reach $\pm 1$ at a band edge, so this is 
perhaps the only way that a new forbidden band can be inserted, 
without disturbing higher bands.

\section{Transmission for odd $N$ systems}
Now we return to the the case of an odd number $N = 2n+1$ of 
half-cells. Since the half-cell already encapsulates the band 
structure of the symmetric double-cell, it might seem that there is 
nothing further to learn from considering the addition of an extra 
factor $W_L$ as in eq. \ref{eq:bip13}. Surprisingly, it changes 
everything. When we form the $M$-matrix from $W_L$, we obtain the 
elements 
 \begin{eqnarray}
M_{11} &=& \cosh \alpha \, \cos \beta - i \cosh \eta \, \sin \beta 
\nonumber \\ 
M_{21} &=& \sinh \alpha \, \cos \beta - i \sinh \eta \, \sin \beta 
\nonumber \\ 
M_{22} &=& M_{11}^* \, ; \quad M_{12} = M_{21}^*\, ; 
\quad {\rm where} \nonumber \\ 
- {\rm Im} M_{21} &=& \half \, \left( \nu u + \frac{g^\pr}{\nu} \right)
\quad \Rightarrow \nonumber \\ 
&& \sinh \eta \equiv \half \, \left(\frac{\nu}{z} - \frac{z}{\nu}\right) ~. 
\label{eq:bip15} 
\end{eqnarray} 

For any odd number of cells, simply replace $\beta \to (2n+1)\beta$ 
leaving the rest unchanged. It is easily checked that det $M = 1$, 
and putting the extra half-cell on the left, rather than the right, 
only reverses the sign of $\alpha$. 

In comparison, for the double-cell one has 
 \begin{eqnarray}
\sinh \mu &\equiv& \half \, \left(\frac{\nu}{Z} - \frac{Z}{\nu}\right) 
= \half \, \left(\frac{\nu}{ e^{\alpha} z} - \frac{z e^\alpha}{\nu}\right)~. 
\label{eq:bip16} 
\end{eqnarray} 
It follows that $\mu =  \eta - \alpha$. 

Using $ 1/t_N = M_{11} $ of eq. \ref{eq:bip15} we have 
 \begin{eqnarray}
\frac{1}{|t_N|^2}  &=& \cosh^2\alpha \, \cos^2 N\beta +  
\cosh^2 \eta \, \sin^2 N\beta \nonumber \\ 
 &=& \cosh^2\alpha  +  [ \cosh^2 \eta - \cosh^2 \alpha] \, \sin^2 N\beta
\nonumber \\ 
 &=& \cosh^2 \eta  -  [ \cosh^2 \eta - \cosh^2 \alpha] \, \cos^2 N\beta
~. 
\label{eq:bip17} 
\end{eqnarray} 
Suppose that $|\eta| > |\alpha|$, making the factor in square 
brackets positive. Then from the second line in eq. \ref{eq:bip17} we 
get a lower bound on the inverse transmission probability, and from 
the third line an upper bound. That is, 
 \begin{eqnarray}
\frac{1}{\cosh^{2} \alpha} &\ge& |t_N|^2 \ge \frac{1}{\cosh^{2} \eta}~. 
\label{eq:bip18} 
\end{eqnarray} 
When $|\eta| < |\alpha|$, the upper and lower bounds are 
exchanged. 
At a point where $\eta = \alpha$, (which implies $\mu = 0$), the 
bounds cross, and $|t_N|^2$ is caught between them. For a symmetric 
half-cell, $\alpha = 0$, so the upper bound on transmission reverts 
to unity, and $\eta$ plays the role of $\mu$ in producing the 
envelope of minimum transmission for the double cell. 

When $|\eta| > |\alpha|$, transmission reaches the upper bound when $N\beta = 
p\pi$, and the lower when $N\beta = (m + 0.5)\pi$. ($p$ and $m$ 
integers.)  In the opposite case, $|\eta| < |\alpha|$, the bounds are 
reversed, and the upper bound is reached at $N\beta = (m + 0.5)\pi$. 
At the ``transparent" point where $|\eta| = |\alpha|$, the 
transmission is pinched between the bounds and takes the value 
$\cosh^{-2} \alpha$. 

\begin{figure}[htb]                     
\begin{center} 
\begin{tabular}{cc} 
a) & \includegraphics[width=8cm]{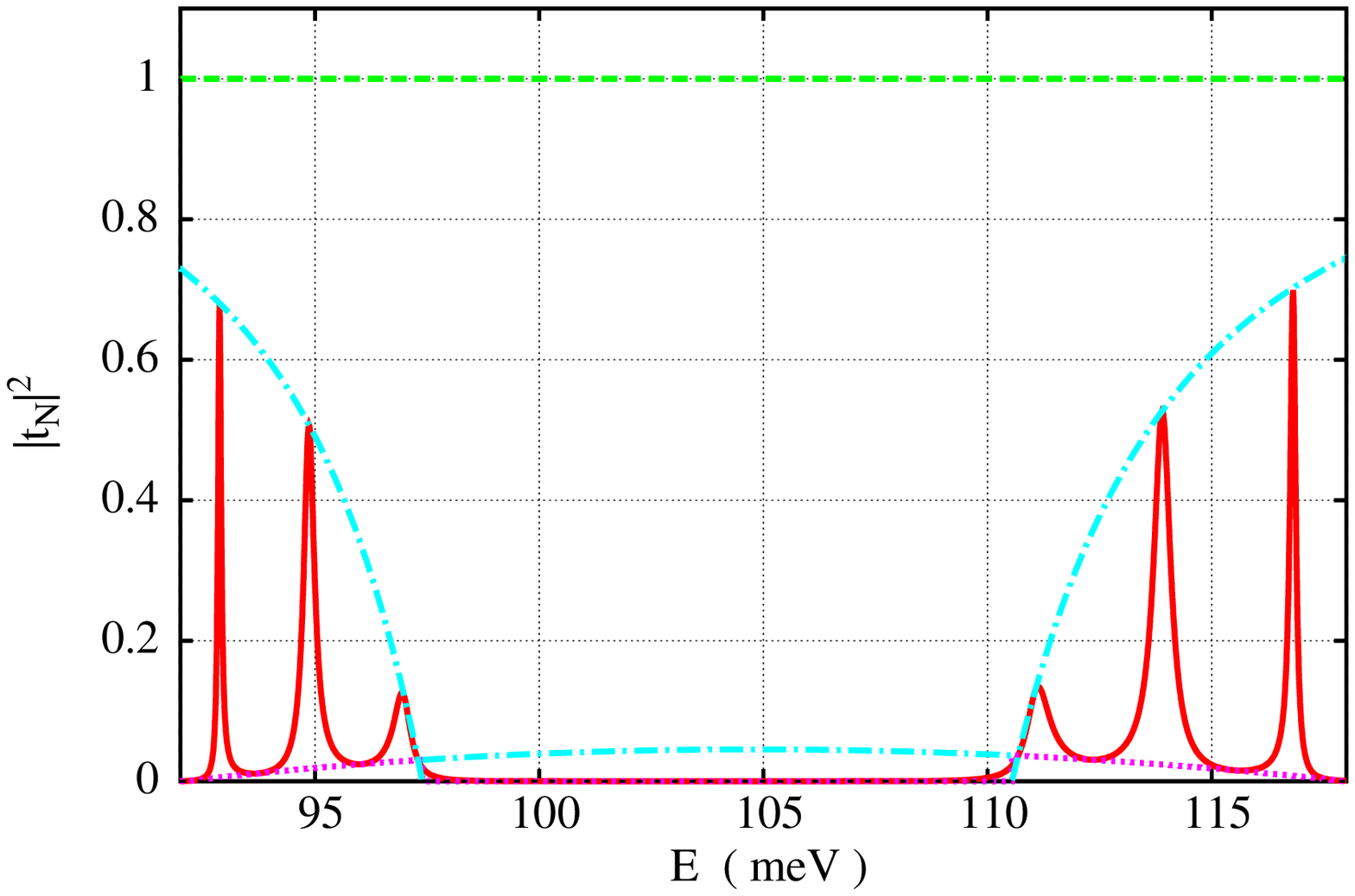} \\ 
b) & \includegraphics[width=8cm]{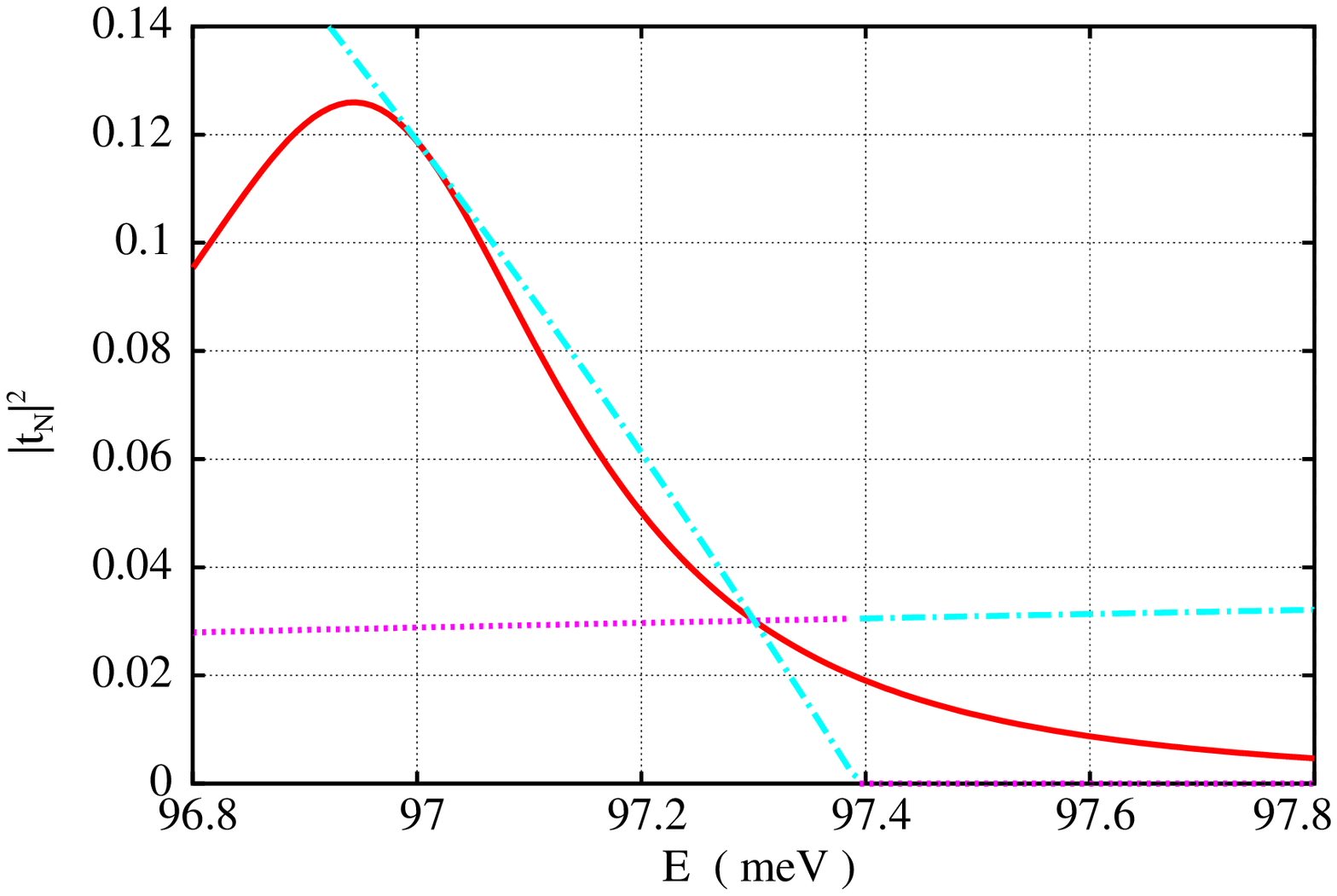} \\ 
\end{tabular}
\end{center}
\caption{(Colour online)  
Transmission of 7 half-cell array as solid (red) line: (a) showing both split bands; 
(b) detail of region around the transparent state. Dash-dot (turquoise) line  
is envelope of maxima; dotted (mauve)line is envelope of minima
up to the transparent point.} 
 \label{fig09} 
\end{figure}

Some results are presented in Fig. \ref{fig09}, for the same 
potential cell used in Fig. \ref{fig02}, appropriate for GaAs/AlGaAs 
superlattices.  To start we take $N=7$ half-cells. Panel (a) shows 
the transmission in both parts of the split bands, with three 
resonances in each. One sees that the envelope of maxima, $1/\cosh^2 
\alpha$ crosses the envelope of minima, $1/\cosh^2 \eta$, at the 
transparent point $\mu=0$. From there to the band edge, they 
exchange their roles as upper/lower bounds. The upper bound 
reduces the maximum transmission by a large amount in comparison to 
the situation in Fig. \ref{fig02}, which is for six half-cells. In 
the forbidden band, it can be shown that 
what was the envelope of mimina becomes an upper 
bound on transmission, while the lower bound is zero. The cross-over at 
the transparent point ensures that these bounds are continuous. 

\begin{figure}[htb]                     
\begin{center} 
\includegraphics[width=8cm]{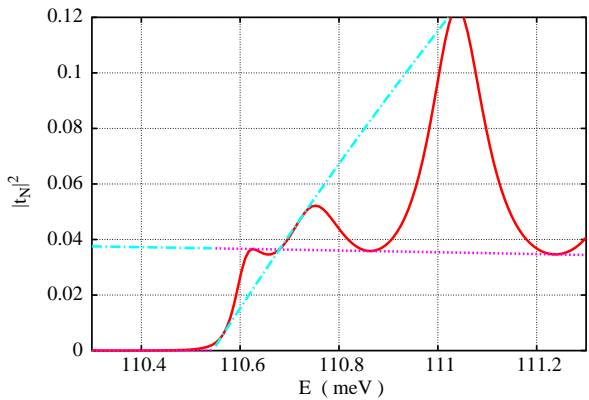} 
\end{center} 
\caption{(Colour online)
Transmission of a 35 half-cell array showing detail in region 
of the transparent state, in the upper split band. Lines as in Fig. 9.}
\label{fig10} 
\end{figure}

Fig. \ref{fig09}(b) shows detail near the transparent point. Where 
the bounds cross, $|t_N|^2$ is pinched between them. To the left of 
the transparent point (which it no longer is!), the maxima occur when 
$N \phi = p \pi$, with integer $p$. From the third line of eq. 
\ref{eq:bip17}, any maxima that occur after the transparent point 
satisfy $N \phi_m = (m+ 0.5)\pi$, with integer $m < n$. Such a situation 
must occur for a large enough $N$, because the energy of the 
transparent point is fixed, while having more cells squeezes 
additional resonances into each allowed band. In Fig. \ref{fig10} we 
show detail of the region near the lower edge of the upper band,  for 
$N=35$ half-cells. Here one of the resonances clearly lies between the 
band edge and the transparent state. 

Innumerable papers have been written on symmetric periodic systems, 
for which the relation eq. \ref{eq:bip01} applies, and their resonances 
always involve perfect transmission. Therefore it is a surprise to 
see how adding a half-cell introduces a non-trivial upper bound on 
transmission, greatly reducing it in the neighbourhood of the 
``transparent" state. 

\section{Conclusion} 
We have studied transmission through biperiodic semiconductor 
superlattices. For an even number of half-cells, asymmetry causes 
each allowed band to split at the Bragg energy where $\cos \phi_h = 
0$. Extending Shockley's line of argument for a generic 
single-barrier cell, we have proved that this induces a transparent 
state which lies in one of the split bands, and very close to the 
band edge. The rule is that the transparent state lies in the lower 
band when the wide well is first in line for incident electrons.  
The transparent state is a resonance which occurs at a fixed energy, 
independent of the number of cells in the array; otherwise 
the transmission follows the well known rule expressed in eq. 
\ref{eq:bip01}. 

For an odd number of half cells, the picture is completely 
different. These systems are asymmetric, so if there is a wide well 
first on the left, there will be a narrow well first from the right. 
The asymmetry due to the additional half-cell causes an envelope of maximum 
transmission to appear, which crosses the envelope of  minimum 
transmission at the ``transparent" point, in both split bands. The 
transmission probability is given by eq. \ref{eq:bip17} and is 
bounded on both sides as in eq. \ref{eq:bip18}. At the transparent 
point $\alpha = \eta$, the bounds cross, and the transmission 
probability is pinched between them. Resonances occurring between the 
transparent state and the band edge satisfy a different rule, 
$N\phi_m = (m + 1/2)\pi$, not the $N\phi_p = p\pi$ which would follow 
from eq. \ref{eq:bip01}.

In a separate article we will discuss these systems in a tight 
binding model, showing what can and cannot be reproduced in that 
approximation. For example, in tight binding, the transparent state 
occurs exactly at a band edge, rather than just inside.

\acknowledgements

We are grateful to NSERC-Canada for support under Discovery Grants RGPIN-3198 
(DWLS), SAPIN-8672 (WvD) and a Summer Research Award through Redeemer 
University College (LWAV); also to DGES-Spain for continued support 
through grants  FIS2004-03156 and FIS2006-10268-C03 (JM).

\appendix 
\section{Transfer matrices} 
For completeness, we write down our conventions for transfer 
matrices in the presence of an energy and position-dependent 
effective mass. The Schr\"odinger equation is 
 \begin{eqnarray}
-m^* \frac{d}{dx} \left[\frac{1}{m^*} \frac{d\psi}{dx} \right] &=& 
 \frac{2m m^*}{\hbar^2} [E - V(x)] \psi(x) \nonumber \\ 
&\equiv&  q^2(x) \, \psi(x)~. 
\label{eq:a01} 
\end{eqnarray} 
The dimensionless effective mass (in units of the free electron mass 
$m$) is $m^*$; $q$ is the wave number inside the potential region, 
and $q(x) \to k$ in the exterior region where both $V(x)$ and $m^*$ 
become constant. (In this paper, the substrate and cap of the 
heterostructure are GaAs, the same as the wells of the superlattice.) 

If $g(x)$ and $u(x)$ are two independent solutions, then 
 \begin{eqnarray}
\frac{\hbar}{m m^*} [g \frac{d u }{dx} - u \frac{d g}{dx}] = {\rm constant} 
\label{eq:a02} 
\end{eqnarray} 
where the constant may be set equal to one by choice of the boundary 
condition at some initial point, say $x=0$. For example, take $g(0) = 
1$, $g^\pr(0) = 0$, $u(0) = 0$ and $[du/dx](0) = m m^*/\hbar$. 

The transfer matrix $W$ relates values of a ``spinor" $\tilde{c}$ 
whose components are $ \psi(x)$ and $(\hbar/m m^*) (d \psi/ dx)$ between two 
points. 
 \begin{eqnarray}
&& \tilde{c}(d) = \begin{pmatrix}\psi(d) \\ 
\big[\frac{\hbar}{m m^*} \frac{d\psi}{dx}\big](d)\end{pmatrix} = 
W_{0,d} \,\, \tilde{c}(0) = \nonumber \\ 
&&  \begin{pmatrix} g(d) & u(d) \\ 
\big[\frac{\hbar}{m m^*} \frac{dg}{dx}\big](d)  & 
\big[\frac{\hbar}{m m^*} \frac{du}{dx}\big](d) \end{pmatrix} \, 
\begin{pmatrix}\psi(0) \\ \big[\frac{\hbar }{m m^*} \frac{d\psi}{dx}\big](0) 
\end{pmatrix}~. 
\label{eq:a03} 
\end{eqnarray}
The Wronskian relation gives det$W = 1$, and in an allowed 
band, Tr$W = 2 \cos \phi$, where $\phi$ is the Bloch phase. 
It is easily shown that 
 \begin{eqnarray}
W^2 &=& 2\cos \phi \, W - I \, , \quad \Rightarrow  \nonumber \\ 
W^N &=& \frac{ \sin( N\phi) }{\sin \phi} \, W - \frac{ \sin (N-1)\phi}{\sin 
\phi} \, I~. 
\label{eq:a04} 
\end{eqnarray} 
The band structure associated with the potential is all contained in 
the $W$-matrix for a single cell. If we agree that the prime symbol 
on the wave function means to take the derivative, and then multiply 
by $\hbar/(m m^*)$, all the above equations reduce to the standard 
form for the case with constant effective mass, in which $m^* \to 1$. 
({\it i.e.} the effective mass can be absorbed into $m$, or one may use an 
average value $\sim 0.071$ and leave it explicit.) 

In discussing transmission properties it is more convenient to 
represent the wave function in terms of plane wave states, normalized 
to unit flux. To the left ($x < x_L $) and right ($ x > x_R$) of the 
potential we write: 
 \begin{eqnarray}
\Psi_L(x) &=& \frac{a_L}{\sqrt{\nu_L}} e^{ik_L(x-x_L)} + \frac{b_L}{\sqrt{\nu_L}}
 e^{-ik_L(x-x_L)} \nonumber \\
\Psi_R(x) &=& \frac{a_R}{\sqrt{\nu_R}} e^{ik_R(x-x_R)} + \frac{b_R}{\sqrt{\nu_R}} 
 e^{-ik_R(x-x_R)} 
\label{eq:a05}
\end{eqnarray}
where the external velocity is $\nu_{L,R} =  {\hbar k_{L,R}} /(m m^*)$, 
with $k_L = \sqrt{ 2m m^* [E - V(x_L)]}/\hbar$, and similarly for 
$k_R$. 
By definition the transfer matrix 
$M$ relates the plane wave coefficients across the system 
 \begin{equation}
\begin{pmatrix}a_L \\ b_L \end{pmatrix} = 
                  M \begin{pmatrix}a_R \\ b_R \end{pmatrix} 
\label{eq:a06}
\end{equation}
Different wave numbers at left and right allow for bias across the potential 
\cite{MSM04}, but here we will not consider that situation further, so 
that $\nu_L =  \nu_R = \nu$.

One can show as in \cite{FPP} that $M$ is related to $W$ by 
 \begin{eqnarray}
M &=& L_L^{-1} W^{-1} L_R \, , \qquad {\rm where} \nonumber \\ 
L &=& \begin{pmatrix} 1/\sqrt{\nu} & 1/\sqrt{\nu} \\ 
           i \sqrt{\nu} & -i \sqrt{\nu} \end{pmatrix} \nonumber \\ 
L^{-1} &=& \half \begin{pmatrix} \sqrt{\nu} & -i/\sqrt{\nu} \\ 
                       \sqrt{\nu} & + i/ \sqrt{\nu} \end{pmatrix}~. 
\label{eq:a07}
\end{eqnarray}
For a double cell as in eq. \ref{eq:bip04}, this leads to 
 \begin{eqnarray}
&& M_{-d,d} = L^{-1} W_{-d,d}^{-1} L = \nonumber \\ 
&&   \begin{pmatrix} \cos \phi -i (\nu ug - u'g'/\nu ) &  i (\nu ug + u'g'/\nu )\\
           -i (\nu ug + u'g'/\nu ) & \cos \phi +i (\nu ug - u'g'/\nu )  \end{pmatrix}
\nonumber \\
\label{eq:a08}
\end{eqnarray} 
and $2 \cos \phi = (ug^\pr + g u^\pr)$. With bias, the $\nu$ in 
the imaginary parts is replaced by $\sqrt{\nu_L \nu_R}$, and the real 
parts become $(\sqrt{\nu_L/\nu_R} \pm \sqrt{\nu_R/\nu_L} )/2$ times 
$\cos \phi$, for the diagonal and off-diagonal elements respectively.

\section{Kard representation in a forbidden band} 

The parameterization given in eqs. \ref{eq:bip12} or \ref{eq:bip13} 
is valid in all allowed bands, if we permit all real values for 
$\beta$ and $z$. In forbidden bands matters are a little more 
complicated. We discuss here the behaviour of the half-cell matrix $W_L$, 
in relation to the bands of the multi-cell system. 

We have in mind a  half-cell of type well-barrier-well, and 
the exterior energy is measured from the well-bottom. At zero energy 
one is in a forbidden zone which we will label (FZ$_0$). All four 
elements of $W_L$ are positive, as are the ratios $g'/g$ and $u'/u$. 
The allowed zone (AZ$_0$) begins when $g'$ becomes negative. The 
parameters $\alpha,\, \beta,\, z$ are all positive in AZ$_0$, with 
$\beta$ in the first quadrant. If we change $\beta \to i\bar{\beta}$ 
and $z \to i \bar{z}$ in FZ$_0$, the correct signs and 
magnitudes are obtained: see line one of eq. \ref{eq:b01}. 

The band AZ$_0$ ends when $g$ also becomes negative, so that the 
log derivatives are both negative, in FZ$_1$. This in turn 
ends when $u'$ becomes negative and the second allowed band AZ$_1$ 
begins. Eventually $g'$ again becomes positive, and FZ$_2$ occurs, 
now with both log-derivatives negative. The signs are summarized in 
Table \ref{table1}, for energies up to 305 meV for a GaAs/AlGaAs SL.
In AZ$_1$, the diagonal elements are negative, which is accommodated 
by placing  $\beta$ in the second quadrant. Since $u$ is still 
positive, $z$ remains positive. Similarly in AZ$_3$, placing 
$\beta$ in the third quadrant replicates the signs. 

In the forbidden zone FZ$_0$, all four elements of $W_L$ are positive. 
In FZ$_1$, both $g$ and $g'$ are negative. Because $\beta = \pi/2$ at 
the band edge, the diagonal elements ($\cos \beta$) are smaller in 
magnitude than the off-diagonal ($\sin \beta$), so $\cos \beta \to 
\sinh \bar{\beta}$ for continuity. The cosh and sinh 
must change places. This is accomplished by $\beta \to \pi/2 + i 
\bar{\beta} $, $\alpha \to \alpha + i\pi/2$, and ${z}$ unchanged. 
The phase $\beta$ increases monotonically through allowed zones, and 
its real part is stationary in forbidden zones. When $\alpha$ 
acquires a phase, $z$ does not. The first few lines of Table 
\ref{table1} summarize these adjustments. 

Similarly, in FZ$_3$, the prescription $\beta \to 3\pi/2 + 
i\bar{\beta}$, $\alpha \to \alpha +i\pi/2$ gives 
the correct signs. There is ample scope to manoeuvre things to fit 
whatever pattern of signs arises, but we doubt that there is any 
general prescription that will fit every possible potential. 

\begin{table}  [htb]
\caption{Signs of elements of the transfer matrix $W_L$, and 
required adjustments of the parameters. (A blank means no 
adjustment.) For real $\beta$, the quadrant is shown.\\ \\} 
\begin{tabular}{|c|c|c|c|c|c|c|c|}
\hline
Band  & $u^\pr$ & $u$ & $g^\pr$ & $g$ & $e^\alpha$ & $z$ & $\beta$ \cr 
\hline
$FZ_0$ & + & + & + & +      &  & $i\bar{z}$ & $i\bar{\beta}$  \cr 
\hline
$AZ_0$ & + & + & $-$ & +    &    &  & I  \cr 
\hline
$FZ_1$ & + & + & $-$ & $-$  & $\alpha +i\pi/2$ & $z$ & $\pi/2 +i\bar{\beta}$ \cr 
\hline
$AZ_1$ & $-$ & + & $-$ & $-$&    &  & II   \cr 
\hline
$FZ_2$ & $-$ & + & + & $-$   &  & $-i\bar{z}$ & $\pi + i\bar{\beta}$  \cr 
\hline
$AZ_2$ & $-$ & $-$ & + & $-$ &  &  & III   \cr 
\hline
$FZ_3$ & + & $-$ & + & $-$   & $\alpha+i \pi/2$ & $-z$ & $3\pi/2 + i\bar{\beta}$  \cr 
\hline
\end{tabular}
\label{table1}
\end{table}

Explicit forms for the first four forbidden bands are 
 \begin{eqnarray}
W_L(FZ_0) &=& 
\begin{pmatrix}e^{\alpha} \cosh \bar{\beta} & (1/\bar{z}) \sinh \bar{\beta} \\ 
           \bar{z} \sinh \bar{\beta} & e^{-\alpha} \cosh \bar{\beta} \end{pmatrix}
= \begin{pmatrix}u^\pr & u \\ g^\pr & g \end{pmatrix} \nonumber \\
W_L(FZ_{1}) &=& 
\begin{pmatrix}e^{\alpha} \sinh \bar{\beta} & (1/\bar{z}) \cosh \bar{\beta} \\
           -\bar{z} \cosh \bar{\beta} & -e^{-\alpha} \sinh \bar{\beta} \end{pmatrix} 
\nonumber \\
W_L(FZ_2) &=& 
\begin{pmatrix}-e^{\alpha} \cosh \bar{\beta} & (1/\bar{z}) \sinh \bar{\beta} \\ 
           \bar{z} \sinh \bar{\beta} & -e^{-\alpha} \cosh \bar{\beta} \end{pmatrix}
\nonumber \\
W_L(FZ_{3}) &=& 
\begin{pmatrix}e^{\alpha} \sinh \bar{\beta} & (-1/\bar{z}) \cosh \bar{\beta} \\ 
           \bar{z} \cosh \bar{\beta} & -e^{-\alpha} \sinh \bar{\beta} \end{pmatrix}~.  
\label{eq:b01} 
\end{eqnarray} 

Transforming from $W_L$ to $M_L$ (for the half-cell), 
 \begin{eqnarray}
M_L &=& \half \begin{pmatrix} 
g+ u^\pr  -i (\nu u - g^\pr/\nu) & g- u^\pr  + i (\nu u + g^\pr/\nu) \\
g- u^\pr  - i (\nu u + g^\pr/\nu) & g+ u^\pr  +i (\nu u - g^\pr/\nu) 
\end{pmatrix}   \nonumber \\ 
\label{eq:b02} 
\end{eqnarray} 
In FZ${_0}$, 
 \begin{eqnarray}
&&\,\, M_{11} = \,\cosh \alpha \, \cosh \bar{\beta} 
                         - i \sinh \bar{\eta} \, \sinh \bar{\beta} 
\nonumber \\ 
&&\,\, M_{21} = -\sinh \alpha \, \cosh \bar{\beta} 
                       - i \cosh \bar{\eta} \, \sinh \bar{\beta} 
\quad {\rm where} \nonumber \\ 
&& \sinh \bar{\eta} = \half \, 
\left(\frac{\nu}{\bar{z}} - \frac{\bar{z}}{\nu}\right)~, \nonumber \\ 
\frac{1}{|t_N|^2} &=& \cosh^2 {\alpha} + \bigg[ \cosh^2 \alpha + 
\sinh^2 \bar{\eta} \bigg] \sinh^2 N\bar{\beta}~. 
\label{eq:b03} 
\end{eqnarray} 
In FZ${_0}$ the upper limit on transmission is given by $1/\cosh^2 
{\alpha}$, as in an allowed band. Also, for large $N$,  $|t_N|^2$ 
approaches zero, as for periodic systems. 
In the next forbidden zone FZ${_1}$, 
 \begin{eqnarray}
 M_{11} &=& -\sinh \alpha \, \sinh \bar{\beta} 
                         - i \cosh \bar{\eta} \, \cosh \bar{\beta} 
\nonumber \\ 
 M_{21} &=& -\cosh \alpha \, \sinh \bar{\beta} 
                       - i \sinh \bar{\eta} \, \cosh \bar{\beta} 
\nonumber \\ 
\frac{1}{|t_N|^2} &=& \cosh^2 \bar{\eta} + \nonumber \\ 
&&\,\, \bigg[ \sinh^2 \alpha + \cosh^2 \bar{\eta} \bigg] \sinh^2 N\bar{\beta}~. 
\label{eq:b04} 
\end{eqnarray} 
The last line of \ref{eq:b04} shows that an upper limit on the transmission 
probability is given by $1/\cosh^2 \bar{\eta}$, but there is no 
lower limit. In Figs. \ref{fig09} and \ref{fig10}, it can be see that 
the upper limit in the forbidden zone is the continuation of 
$\cosh^{-2} \eta$ from the adjacent allowed zone. 

The roles of $\alpha$ and $\bar{\eta}$ are exchanged 
between eqs. \ref{eq:b03} and \ref{eq:b04}. This pattern applies to 
all FZ with even and odd indexed subscripts. 

In FZ$_2$, the expressions for $M_{11}$ and $M_{21}$ are the complex 
conjugates of those in eq. \ref{eq:b03}, and in FZ$_3$ they are 
complex conjugates of those in eq. \ref{eq:b04}. This pattern appears 
to persist in going to higher bands, but may depend on the detailed 
form of the potential. The forbidden zones with even and odd labels 
have different patterns, due to the need to exchange the roles of 
$\cosh \bar{\beta}$ and $\sinh \bar{\beta}$ in the latter case.

The main virtue of the Kard parameterization is that, in allowed 
bands, it makes the relations between the single and multiple cell 
systems very obvious. Kard is less convenient in forbidden bands, 
but with sufficient care it offers the same advantages.

\end{document}